\documentclass[a4paper]{jpconf}
\usepackage{graphicx}
\begin{document}
\title{The ALICE analysis train system}

\author{Markus Zimmermann for the ALICE collaboration}

\address{CERN, 1211 Geneva 23, Switzerland}

\ead{m.zimmermann@cern.ch}

\begin{abstract}
In the ALICE experiment hundreds of users are analyzing big datasets on a Grid system. High throughput and short turn-around times are achieved by a centralized system called the ’LEGO’ trains. This system combines analysis from different users in so-called analysis trains which are then executed within the same Grid jobs thereby reducing the number of times the data needs to be read from the storage systems. The centralized trains improve the performance, the usability for users and the bookkeeping in comparison to single user analysis. \\
The train system builds upon the already existing ALICE tools, i.e. the analysis framework as well as the Grid submission and monitoring infrastructure. The entry point to the train system is a web interface which is used to configure the analysis and the desired datasets as well as to test and submit the train. Several measures have been implemented to reduce the time a train needs to finish and to increase the CPU efficiency. 
\end{abstract}

\section{Introduction}
The ALICE collaboration is recording around 10 PBs every year. This amount of data is stored on different storage elements all over the world.
To analyze a single dataset, on the order of $10^5$ files have to be read where those files containing data on the order of 50 TB. The creation and submission of the analysis jobs to the Grid can be done with the ALICE analysis framework [1]. These jobs can be monitored and resubmitted using a monitoring service (MonALISA [2]).\\
On top of these existing frameworks the LEGO (Lightweight Environment for Grid Operators) train system was created. It was developed to increase the CPU efficiency of the analysis jobs and thus to get more user analysis done with the same amount of computing resources. This is done by collecting multiple analysis tasks which should analyze the same dataset and by running them within the same analysis job. Following this procedure the data is read once, and then it is used for multiple analysis tasks. In comparison to that if the different user analysis would be done individually the data would be read for each analysis separately. This way more time would be used for loading the data and the CPU efficiency would be lower.\\
Another advantage of the LEGO trains is to hide the Grid complexity from the users. They just have to define their code on a web page which also provides the analysis results as soon as it is available. In between, the system computes the results fully transparent for the user. This process is fully automatized so that no one needs to supervise the jobs which are running on the Grid. Further improvements are implemented in the computing process to use the available resources as efficient as possible. \\
The train system contains several trains divided by physics working groups (jet analysis, particle correlations, ...) and different kind of datasets (pp-, pPb- or PbPb-collisions, data or Monte Carlo). The analysis of a dataset by a train is called a train run.

\section{Running a LEGO train}
\begin{figure}[h]
\includegraphics[width=0.80\columnwidth]{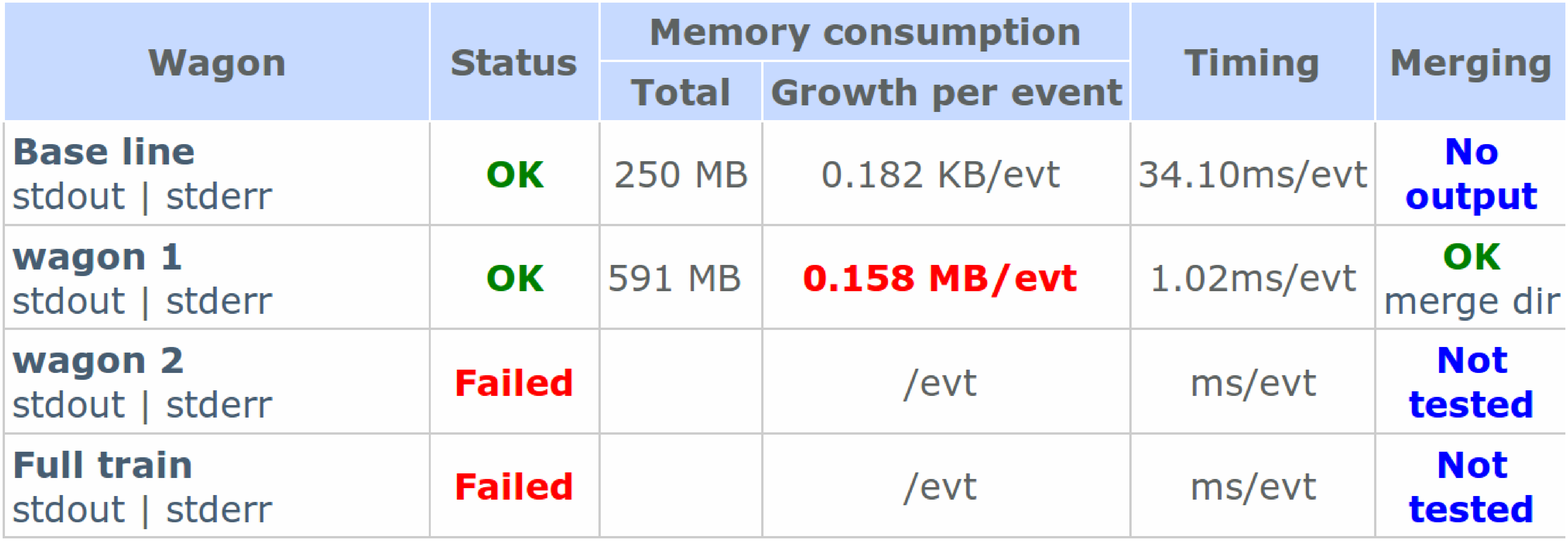}\hspace{2pc}%
\caption{\label{fig:trainTest}Example test of a train run. The Base line does not contain any user code, the wagons contain just their own code and the Full train contains all the wagon codes.}
\end{figure}

Before running in a train a user has to submit the analysis code to a repository. Then the configuration of the code has to be defined on a web page in a so-called train wagon. \\
The trains are managed by train operators who define the datasets and configure the jobs of the train runs. On a regular time basis, which is usually once per day, the trains are started. To do so an operator composes a train run out of a set of wagons and a dataset. The train run is tested with an automatized procedure and only if the test is successful the train run can start on the Grid.\\
An example of a train test is shown in figure \ref{fig:trainTest}. The Base line tests the configuration of the train and does not contain any user code. The Full train contains all the train wagons. In between all wagons are tested individually. The test reports several measurements. The first one is the status if the test is finishing normal (OK) or if there is an exception in the code (Failed). The memory consumption is divided into the total consumption of the test and the growth per event. The later number indicates a memory leak if it is getting too big. To show this to the operator the number is highlighted in red if a memory leak is suspected. The timing information shows the used time per event. This allows a rough estimate of the train duration. In the last column the result of the merging test is reported.
\begin{figure}[h]
\includegraphics[width=0.80\columnwidth]{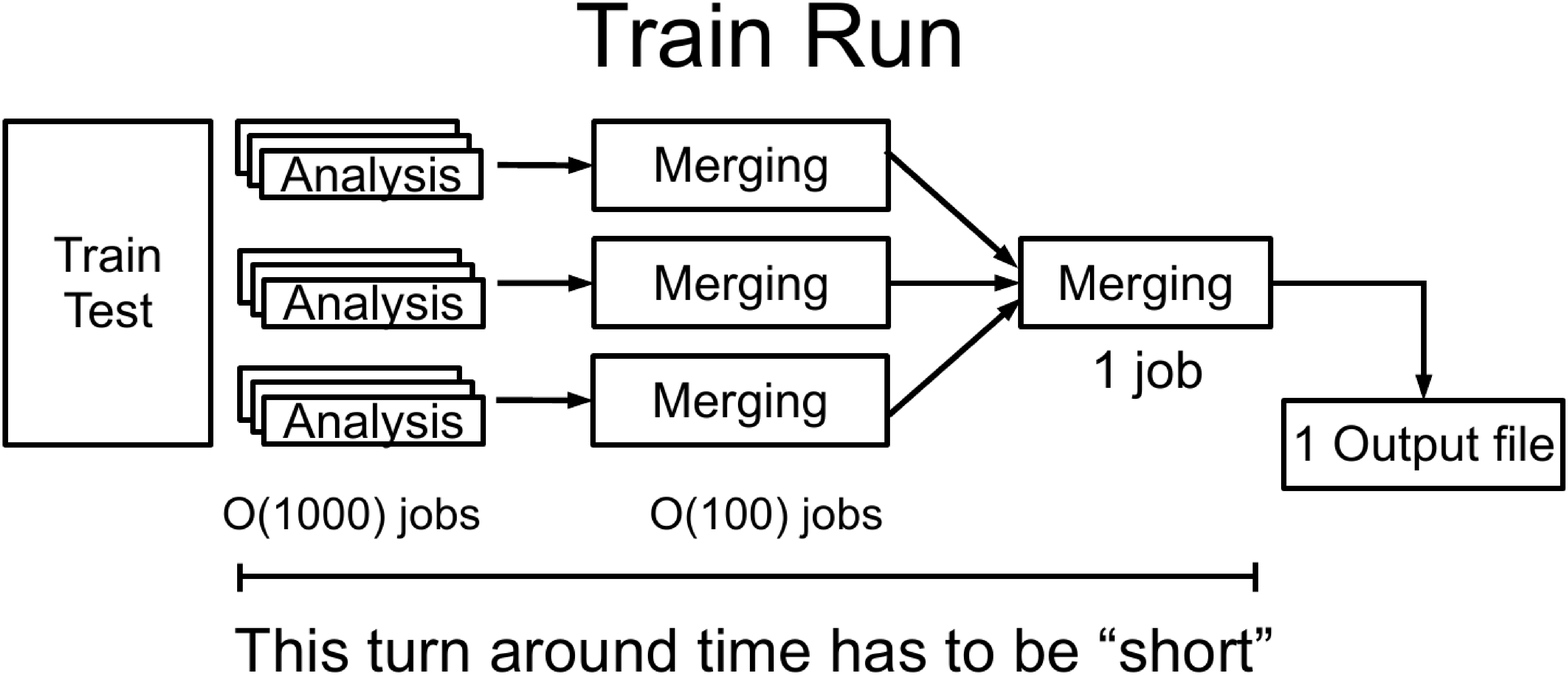}\hspace{2pc}%
\caption{\label{fig:trainRun}Procedure of a train run. After the train test the analysis jobs run. Then the results are merged into a single output file. The time to complete this procedure is called the turn around time.}
\end{figure}%
\\If there is an issue in the test, the operator has to fix this or exclude the failing wagon(s). Afterwards the test has to be repeated. If the test is successful the train run can start. The chain of jobs run on the Grid is shown in figure \ref{fig:trainRun}. After the test, the analysis jobs are started. These jobs are grouped into sets which are merged as soon as all the jobs finished. The results of the merging jobs are merged again in a last merging job. The output of this last job is the result of the train run and is linked on the LEGO train web page. The time from starting the first analysis jobs until the merging is finished is called the turn around time. This is the time the user has to wait for the output and it should be as short as possible.\\
The configuration files of the train run and the final result can be downloaded from the web page by everyone inside ALICE. This way all the train runs are documented in the same way and it is easy to hand results to someone else within the experiment.

\section{Statistics}
\begin{figure}[h]
\includegraphics[width=\columnwidth]{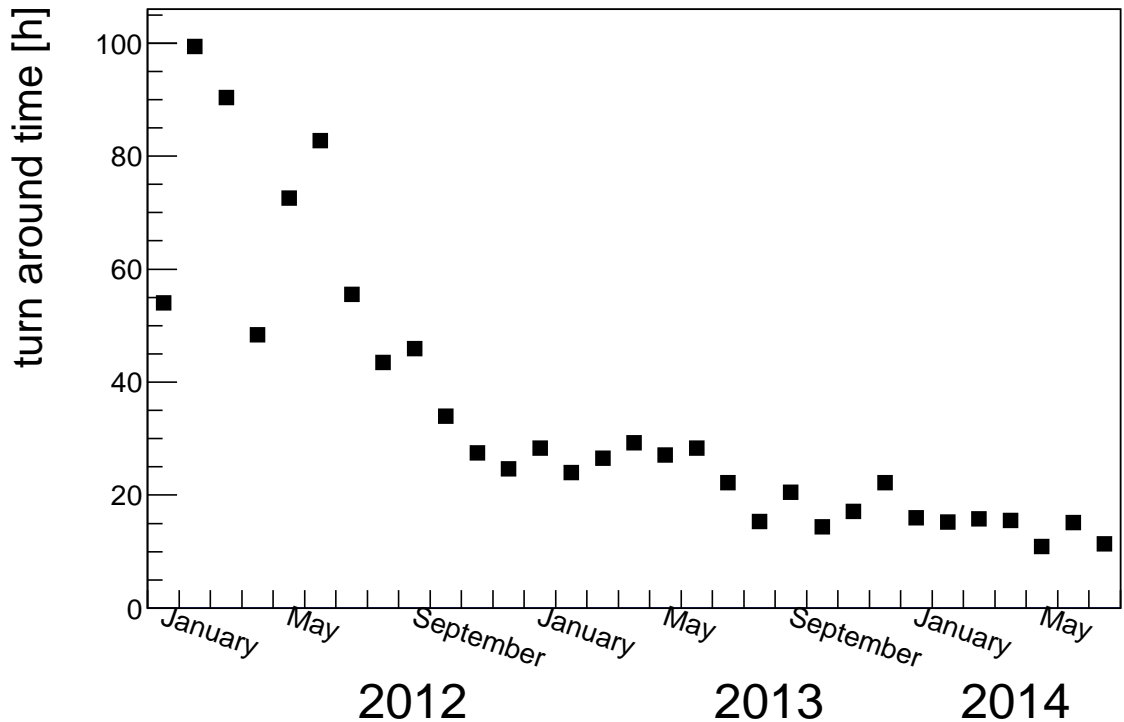}\hspace{2pc}%
\caption{\label{fig:avgTurnAround}Average turn around time of the trains per month.}
\end{figure}

In table \ref{tab:usageStatistics} the usage statistics of the LEGO train system are shown. The system was set up in the beginning of 2012 and since then has continuously run. For 2014, the numbers are extrapolated from the measured values within the first 6 months of the year.\\
The line 'part of the user analysis done with the analysis trains' represents the total amount of computing time on the grid used by the trains divided by the total amount of computing time on the grid used for ALICE analysis. The value is increasing from 27\% in 2012 to 70\% in 2014. In the same time the number of users increased from 60 to 188. This shows that the LEGO trains are well accepted among the users and that they are now used for most of the analysis in ALICE. \\
For the users, the average turn around time is an important value. This value shows how long they have to wait on average to get the train results after the train has been submitted to the Grid. In figure \ref{fig:avgTurnAround}, the development of the average turn around time per month is shown. It improves from 3 to 4 days at the beginning of 2012 to 14 hours in 2014. The reduction in July 2013 is achieved by putting the improvements discussed in section \ref{kap:convince} into production.
\begin{center}
\begin{table}[h]
\centering
\caption{\label{tab:usageStatistics}Usage statistics of the LEGO trains.} 
\begin{tabular}{@{}l*{15}{r}}
\br
 &2012 & 2013 & 2014 \\ 
 &     &      &(extrapolated)\\
\mr
Users & 60 & 127 & 188\\
Trains & 42 & 69 & 79 \\
Train runs & 1537 & 4794 & 7447 \\
Number of jobs & 12M & 26M & 34M \\
Train wagons/run & 14.9 & 10.1 & 8.9 \\
\vspace{-0.1cm}
Part of the user & & & \\
\vspace{-0.1cm}
analysis done & 27\% & 57\% & 70\% \\
with the trains & & & \\
Processed data & --- & 75PB & 130PB \\
Turn around time & 49h & 22h & 14h \\
\br
\end{tabular}
\end{table}
\end{center}

\newpage

\section{How to convince users to use the LEGO trains} \label{kap:convince}
Because the train system is collecting multiple analysis to run them together, the system works better when more people are using it. This section explains how the users were convinced to use the system.\\
The biggest advantage for the users is that they save time. The setup of a wagon is much easier and faster than submitting all the jobs for an analysis. Additionally, the resubmission of failed jobs, the submission of the merging jobs and the bookkeeping are done automatically by the system. As soon as a train finishes the user is notified by an email. Also new users can learn faster how to run an analysis via the train system than running it directly on the Grid.\\
The definition of the datasets is done by operators. This means that some datasets can already be divided into good and semi-good runs by a previous quality analysis. So the user just needs to choose on which dataset to run instead of defining it.\\
Some job parameters are predefined by the system or they are defined by the operators. This leads to a stable job management which has less failed jobs compared to user managed analysis. Additionally, trains make it easier for the operators and the Grid support personnel to find bugs in the jobs so that the users get faster support.

\section{Improvements}
One of the crucial parameters for the users is the turn around time of the trains. The objective of the improvements was to achieve a turn around time of one night. This way the user can define their analysis task in the evening and then the train runs over night. On the next morning, the user can continue with their work. \\
In this section some of the most effective improvements are discussed.

\subsection{Job submission}
The ALICE computing model schedules the analysis jobs always to the computing sites which have the data on a local storage element (SE) to minimize the loading time. This can lead to the problem that some jobs have to wait for a long time because all job slots at the computing sites with a local copy of a certain file might be occupied. To solve this, the local data requirement is dropped for the train jobs as soon as 90\% of the jobs are successfully finished. Then the remaining jobs can be analyzed on any computing site and the data is read remotely. Although this increases the time needed to read the data, it decreases the time until the last jobs are started. This speeds up the last jobs of the train and thus also the full train run itself. This is done at the cost of the CPU efficiency.\\

\subsection{Dataset consolidation}
One job can only analyze multiple input files if they are saved at the same local SE. Most of the datasets (especially from MC productions) are spread over a lot of different SEs. So they end up having a lot of jobs which are processing just 1 input file. In addition to the physics analysis, each job produces some overhead for producing and starting the job. \\
The number of input files in a job can be increased by consolidating the dataset. This is done by copying some of the files to another SE so that more files can be executed within the same job. This is only done with the most used datasets. For some datasets, the consolidation reduces the number of jobs up to about 90\%. Running fewer jobs means less overhead for the train run.

\subsection{Run time of the jobs}
The introduction of a monitoring system to see the run time of the analysis jobs showed that most of the jobs are finishing within a quite short time. For most of the train runs, more than 95\% of the jobs finish in half the time it takes the longest running job to finish. This offers a new way to speed up the train turn around time. The last 2\% of the active jobs are killed and then the merging is started directly. This way some statistics is traded for a faster turn around time. This can only be done if the distribution of killed jobs is not introducing a bias on the data analysis.

\section{Conclusions}
The LEGO train system is a workflow for the organized analysis in ALICE. The system is commonly used and consumes 70\% of the CPU time dedicated to user analysis. It hides the Grid complexity from the users and does the job submission, resubmission and bookkeeping automatically. The system saves computing resources by combining several user tasks in one analysis job. This results in a better CPU efficiency by loading the data less often.\\
By using the system, the Grid support personnel can find bugs easier and thus give faster support. This saves a lot of time for the users and the Grid support personnel compared to a situation in which all users do their own individual user analysis.
\\
\\
\\
\textbf{References}\\
\begin{description}
 \item [{[1]}] Gheata A 2008 ALICE analysis framework \textit{Proc. XII Int. Workshop on Adv. Comp. and Analysis Techniques in Phys. Research (Erice) PoS(ACAT08)} p 028
 \item [{[2]}] I. Legrand et al., “MonALISA: An agent based, dynamic service system to monitor, control and optimize grid based applications”, \textit{Computer Physics Communication}, vol. 180, no. 12, pp. 2472 – 2498, 2004
\end{description}

\end{document}